\documentclass[letterpaper, 10 pt, conference]{ieeeconf}  % Comment this line out if you need a4paper
\usepackage{graphicx}
\usepackage{graphics} % for pdf, bitmapped graphics files
\usepackage{epsfig} % for postscript graphics files
\usepackage{mathptmx} % assumes new font selection scheme installed
\usepackage{times} % assumes new font selection scheme installed
\usepackage{amsmath} % assumes amsmath package installed
\usepackage{amssymb}  % assumes amsmath package installed
\usepackage{hyperref}
\usepackage{multirow}
\usepackage[table]{xcolor}% http://ctan.org/pkg/xcolor
\usepackage{cite}
\usepackage{longtable}
\usepackage{booktabs}
\usepackage{wrapfig,lipsum}
\usepackage{xtab}
\IEEEoverridecommandlockouts                              % This command is only needed if 
\title{\LARGE \bf React to This! How Humans Challenge Interactive Agents \\using Nonverbal Behaviors}

\author{Chuxuan Zhang$^{1}$,  Bermet Burkanova$^{1}$, Lawrence H. Kim$^{1}$, Lauren Yip$^{1}$, Ugo Cupcic$^{2}$,\\ Stéphane Lallée$^{2}$, and Angelica Lim$^{1}$% 
\thanks{$^{1}$C. Zhang,  B. Burkanova, L. Kim, L. Yip, and A. Lim are with the School of Computing Science,
        Simon Fraser University, 8888 University Dr., Burnaby, Canada
        {\tt\small \{chuxuan\_zhang, bermet\_burkanova, lawkim, lauren\_yip,angelica\}@sfu.ca}}%
\thanks{$^{2}$ U. Cupcic and S. Lallée are with Spoon AI, 4 rue de la Bourse, 75002 Paris, France
        {\tt\small  \{ugo, stephane.lallee\}@spoon.ai}}%
}

\begin{document}

\maketitle
\thispagestyle{empty}
\pagestyle{empty}

% \begin{figure*}[ht]
%     \centering
%     \includegraphics[width = 0.9\textwidth]{Figures/characters.png}
%     \label{fig:enter-label}
%     \caption{Participants were asked to test six interactive virtual characters physically, emotionally and socially}
% \end{figure*}
How do people use their faces and bodies to test the interactive abilities of a robot? Making lively, believable agents is often seen as a goal for robots and virtual agents but believability can easily break down. In this Wizard-of-Oz (WoZ) study, we observed 1169 nonverbal interactions between 20 participants and 6 types of agents. We collected the nonverbal behaviors participants used to challenge the characters physically, emotionally, and socially. The participants interacted freely with humanoid and non-humanoid forms: a robot, a human, a penguin, a pufferfish, a banana, and a toilet. We present a human behavior codebook of 188 unique nonverbal behaviors used by humans to test the virtual characters. The insights and design strategies drawn from video observations aim to help build more interaction-aware and believable robots, especially when humans push them to their limits.

% believability of interactive virtual characters 
% but breaks down in open environments
% as humans test the interaction capabilities of the agents
% what should characters react to

\section{Introduction}
Imagine walking into a theme park and seeing an animatronic character, smiling at you. To test if it can see you, you immediately shift from left to right, and its gaze seems to follow you everywhere you go. You wave hello, and surprisingly, it waves back. You make a silly face to see if it will react to more complicated gestures. But this time the spell is broken -- it doesn't respond to you at all!

The concept of believability has long been explored in animation and artificial intelligence, from video games~\cite{curtis2022toward} and human-robot interaction (HRI)~\cite{ribeiro2012illusion} to interactive virtual agents~\cite{pelachaud2003computational}. One important aspect of believability is interaction awareness~\cite{bogdanovych2016makes}. Interaction awareness is defined as the ability of an agent that is “to perceive important structural and/or dynamic aspects of an interaction that it observes or that it is itself engaged in”~\cite{dautenhahn2002embodied}. Non-verbal interactions with artificial agents, like the creature described above, remain challenging to produce convincingly and appropriately~\cite{wang2021examining}.

%Paragraph 2: Limitations
Indeed, systems that produce believable, lively non-verbal interactions are rare~\cite{curtis2022toward} and believability can break down quickly in free interaction environments. We suggest that a cause for this breakdown is that the set of non-verbal behaviors (e.g. facial, and body gestures) that people use to test an agent's capabilities is not clear. In addition, how people behave may depend on the character's appearance and expected social and physical capabilities~\cite{10.1145/3477322}; for example, a robot with no hands may be less likely to be offered a handshake. In essence, in addition to a wave or smile, what other interactive behaviors should a robot be prepared to react \emph{to}? 

%Paragraph 3: What we did do and contribute
The main contributions of this paper are:
\begin{enumerate}
 \item A study of nonverbal behaviors that people use to test a character physically, emotionally, and socially
 \item A human behavior codebook of 188 action classes that researchers can consider for recognition in HRI
 \item Design insights related to a character's interactive affordance
    % \item \textbf{Providing design insights into how the morphology of a virtual character affects human interactive behaviors.} We explore 6 characters' morphologies and the related sets of behaviors people use to interact with them. We then discuss the concept of expectation in the context of character interaction.
\end{enumerate}

\begin{figure}[t]
    % \centering
    \includegraphics[width = 0.49 \textwidth]{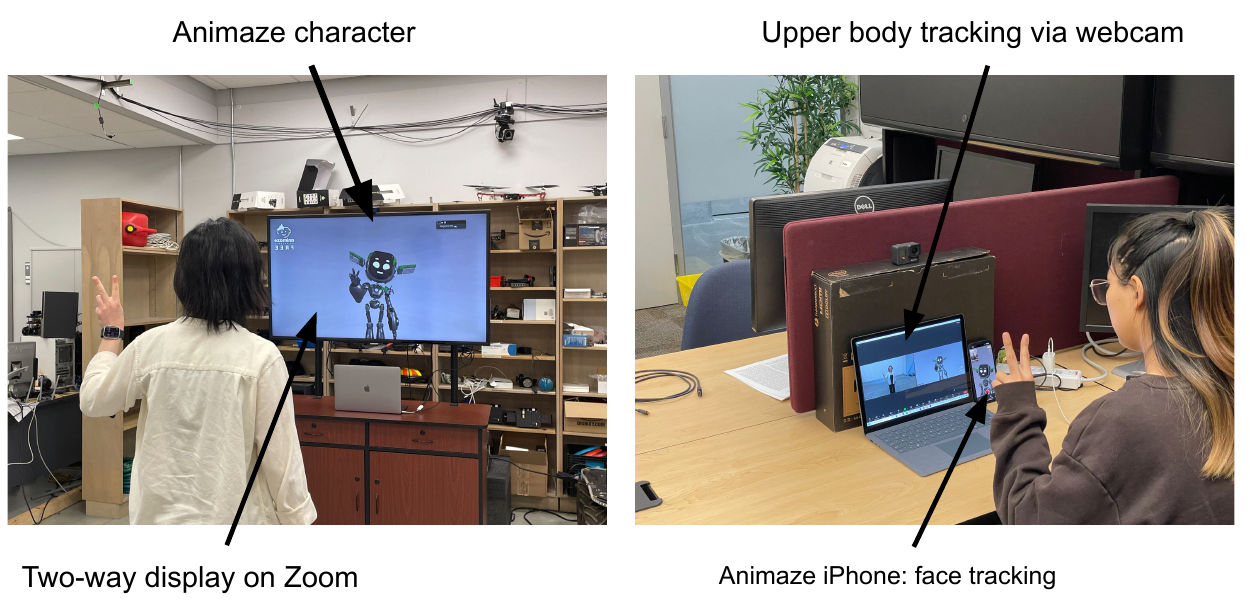}
    
    \caption{Physical setup of our study: participant interacting with virtual character (left), and teleoperator using face and upper body tracking to control the virtual character (right).}
  
    \label{fig:setup}    
    % \Description{Two photos capture the physical setup of the study. The photo on the left shows a participant interacting with a virtual character displayed on a smart screen. The photo on the right shows the teleoperator remotely controlling the virtual character from another room with an iPhone face tracker, and the webcam on a laptop.}
\end{figure}

% We asked participants to test what characters could and couldn't do physically, emotionally and socially. We recorded 1170 behaviors, which we then filtered into a codebook of 137 unique actions, classified into 48 psychosocial behaviors categories which designers can consider towards building believable experiences. In addition, we explore 6 characters' morphologies and the related sets of behaviors people use to interact with them. We then discuss the concept of affordance in the context of character interaction.\\
%As a practical step, this paper provides a study of common non-verbal behaviors that test character interaction awareness. In this study, we asked participants to test what the characters could and couldn't do: physically, socially and emotionally. We recorded 1099 behaviors and categorized them into a codebook of 100 different triggers that designers can consider when designing believable experiences. In addition, we explore 6 characters' morphologies and their related sets of challenge behaviors, then discuss character interaction affordance.

% , hereafter referred to as "challenge behaviors"

\section{Related Work}
%Short paragraph 1: A sentence about related areas
This research builds upon prior work on testing for awareness along with nonverbal and affective behaviors that arise in interactions with artificial agents. We specifically focus on a subset of socially interactive agents~\cite{10.1145/3477322} including social robots and interactive virtual agents (IVAs) that rely on visual sensing devices such as cameras. We refer to these agents in this paper as (interactive) characters.
%, which specifically rely on visual sensing devices such as a camera, rather than a mouse or traditional gamepad controller

%Section 1: Testing for awareness
%\subsection{Animacy and Interaction Awareness}The concept of \emph{animacy} was described in 1929 by Jean Piaget in his study of children and their "tendency to regard objects as living and endowed with will"~\cite{piaget1926representation}. Piaget’s framework suggests that movement and intentional behavior are important for the perception of animacy~\cite{bartneck2009measurement}, and more recent studies show that even infants as young as 9 months old  differentiate animate from inanimate robots on the basis of motion cues \cite{POULINDUBOIS199619}. 

%Indeed, hidden cameras following a humanoid robot in a shopping mall in Japan showed that children performed behaviours such as kicking and covering the robot's eyes. While such actions could be viewed as bullying, when asked why they did this, the children replied that they were ``curious"~\cite{nomura2015children}. This is an illustration of how humans may test the animacy and abilities of a robot, including its interaction awareness~\cite{bogdanovych2016makes}.

\emph{Interaction awareness} involves the ability to perceive dynamic aspects of an interaction~\cite{dautenhahn2002embodied}. Specifically, Dautenhahn et al. suggest that an ``important ability of an interaction-aware agent is to \emph{track, identify, and interpret visual interactive behavior}"~\cite{dautenhahn2002embodied}, along the continuum of interaction formality~\cite{hutchby2008conversation}. This includes informal interactions such as play, semi-formal interactions such as greetings, and very formal interactions such as scripted law proceedings. As an example, Aldebaran Robotics\footnote{https://www.aldebaran.com/} proposed the Basic Awareness module on their NAO and Pepper robots, which includes tracking detected humans and looking in the direction of stimuli such as movement or sound, towards the illusion of life. 

%In human-robot interaction (HRI) studies, numerous studies have used the Animacy scale from the Godspeed Questionnaire Series \cite{bartneck2009measurement} which measures the perceived animacy of an agent. In these studies, it is common to change the robot's interactive behaviors (e.g. gaze, gesture) then ask participants to rate the agent on 6 scales, including dead/alive, stagnant/lively, mechanical/organic, artificial/lifelike, inert/interactive, apathetic/responsive~\cite{bedia2014quantifying}.  Another similar scale~\cite{jannai2023human} uses four items on a 10-point Likert scale: lifelike, machine-like, interactive, and responsive. Towards improving ratings on these scales, in this paper, we focus on Dautenhahn~\cite{dautenhahn2002embodied}'s suggestion to track, identify, and interpret a human interactor's visual interactive behaviours.
%To the best of our knowledge, a measure of a robot's "robot-human" interaction awareness is still to be created, and in this paper, our quantitative analyses primarily use the well-defined Godspeed Animacy scale.

%Section 2: Non-verbal behavior
%\subsection{Tracking, Identifying and Interpreting Human Visual Behaviors in Interactions}
\emph{What do humans do when faced with an interactive agent?} Human non-verbal and affective expressions have been studied for many years, resulting in numerous hand and body gestures \cite{zhang2018egogesture,luo2020arbee} and facial expression datasets \cite{kollias2023abaw,liu2021imigue} capturing expressions corresponding to emotional labels such as anger, happiness, excitement, sadness, frustration, fear, and surprise~\cite{busso2008iemocap}. Decades of psychological research have studied human-human interactions, to understand body gestures ~\cite{busso2008iemocap, volkova2014mpi, matsumoto2013cultural} including emblems~\cite{ottenheimer2018anthropology, matsumoto2013emblematic}, i.e. gestures that can replace speech, such as head motions for yes, no, or a shrug indicating I don't know. Interactive virtual agents such as Greta~\cite{10.5555/1558109.1558314}, SimSensei ~\cite{10.5555/2615731.2617415} and M-PATH \cite{yalccin2020m} aim to respond to human behaviors by specifically tracking features such as gaze, facial expressions and body movement to estimate the mental and emotional state of a user to empathize or change its verbal responses. The social signals created for these agents constitute a collection of behaviors relevant to this paper.

\emph{Data-driven behavior databases.}
In the human visual behavior databases mentioned above, study participants express or annotate behaviors assuming \emph{a priori} class labels. But what if the labels are unknown? Another, more naturalistic, paradigm is to allow expressions to emerge from interactions, and then perform post-hoc labeling. As an example, in the Tower Game~\cite{salter2015tower} experiment, participants were asked to build a tower out of blocks and were not allowed to talk, resulting in only non-verbal interactions and expressions. Gestures such as kiss and peek-a-boo have also been uncovered in data-driven analyses of physical and social interactions specifically with infants, and have limited verbal interaction capabilities~\cite{colgan2006analysis}. One of the most related studies to our work explores abuse of robots by children, analyzing 12 hours of behavior in a shopping mall and describing a handful of bullying behaviors such as grabbing, pulling, blocking, and shoving robots \cite{kanda2020}.

Overall, despite the many studies studying nonverbal social and affective behaviors between humans and artificial agents, a) many of them focus on modeling the agent's behavior rather than the human's, and b) datasets annotating human interactive behavior still lack information regarding how people challenge virtual agent and machine interactive abilities.
%, and c) how morphology of the virtual character affects what people do remains an open question. In sum
In order to design interactive agents that react appropriately to human behavior, it is imperative to first understand what exactly the agents should react \emph{to}.

% Add Contributions summary

\begin{figure}[t]
    % \centering
    \includegraphics[width = 0.47 \textwidth]{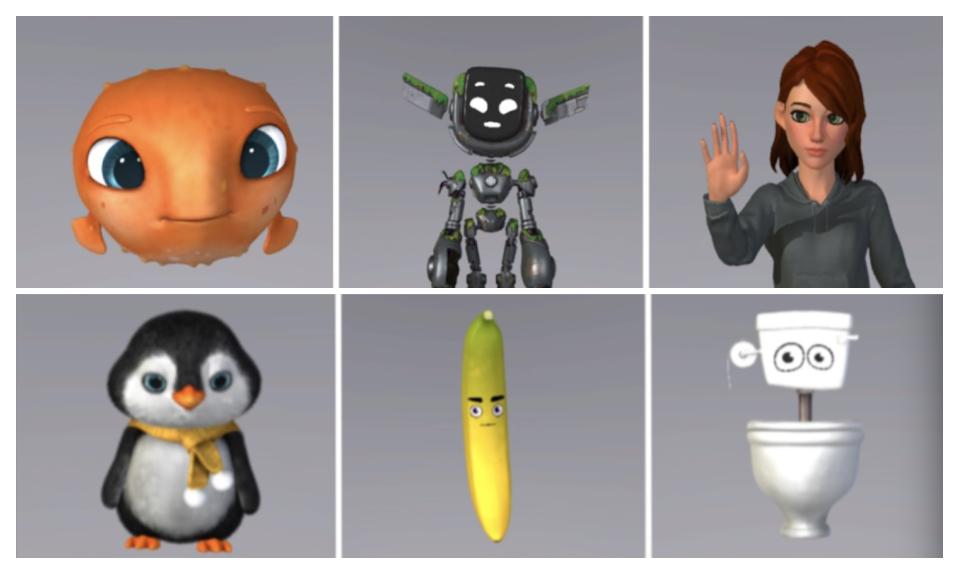}
   
    \caption{Participants were asked to test 6 interactive virtual characters physically, emotionally, and socially.}

    \label{fig:character}    
    % \Description{Two photos capture the physical setup of the study. The photo on the left shows a participant interacting with a virtual character displayed on a smart screen. The photo on the right shows the teleoperator remotely controlling the virtual character from another room with an iPhone face tracker, and the webcam on a laptop.}
\end{figure}

\section{Methodology}
\subsection{Research Questions}
Our long-term goal is to improve a character's interaction awareness. In this study, we address a specific knowledge gap in the context of a \emph{one-to-one, non-verbal interaction with an interactive agent} (Fig.~\ref{fig:setup}, left). We investigate the research question (RQ): \textbf{What do people do to test a character physically,  emotionally, and socially?}

%\item{RQ 2: How do these behaviors change depending on the character's morphology?}

%\item{RQ 3:} What insights can be gained by analyzing the patterns of non-verbal interactions and interview data?

\subsection{Setup and Materials}

We prepared a Wizard-of-Oz(WoZ) experimental setup using three off-the-shelf software packages to create interactions between human participants and virtual characters (Fig.~\ref{fig:setup}). We used Animaze\footnote{https://www.animaze.us/}, a software that provides ready-to-use animated characters of various morphologies, accompanied by its built-in iPhone face tracking module and Webcam Motion Capture\footnote{https://webcammotioncapture.info/} to track the teleoperator's upper body including head, shoulders, and finger movements. 
Zoom\footnote{https://zoom.us/} was used for this study to allow the teleoperator to see the participants' behaviors and vice versa. On the participant side, the virtual character was displayed on a large TV screen (55 inches) oriented horizontally, slightly above eye level. Participants were asked to stand at a marked location approximately 1.5m from the display.

% The full software pipeline is in Fig. \ref{fig:software}.

\subsection{Pilot Study}

Five participants were recruited to refine the study design. In this pilot study, we used the physical setup depicted in Fig.~\ref{fig:setup}. Each of the 5 participants interacted with 11 different agents. The task given to the participants interacting with the virtual character was: ``These are free interactions, and you can do whatever you want." To keep all participants in the main study receiving the same treatment, we designated a single teleoperator and wrote a concise interaction guideline for the teleoperator. The teleoperator was told not to intentionally start any interaction, but only react to participants' actions, so that we could observe interactions that are only initiated by participants. 
% no space for appendix

\label{subsubsec: behavioral categories}
We observed 3 major behavioral categories in the pilot: 1) \emph{Posture, proxemics and physical contact}: Pilot participants sometimes adjusted their \emph{posture} \cite{howorth1946dynamic} (e.g. tilted their head), and moved their bodies subtly (e.g. wiggled body) during the interaction. Participants also changed their relative distance to the character, for example by walking towards the character, i.e. \emph{proxemics}\cite{hall1963system}. Finally, we encountered faked \emph{physical contact} such as poking the character, similar to haptic interactions without touching the screen. 2) \emph{Affect Displays}: Some pilot participants showed frowning eyebrows when seeing the characters' unexpected reactions, and smiled when being amused by the characters. 3) \emph{Emblematic Gestures}: We saw that some pilot participants tried to compliment the characters by giving thumbs up and raising one hand to ask for a high-five. 

As a result of the pilot study, we made 3 changes to the study design. First, we reduced the number of characters to interact with from 11 to 6 due to the boredom and tiredness reported by the pilot participants. The six characters (Fig. \ref{fig:character}) were chosen to cover a broad range of morphologies: 1) a pair of humanoid characters (a robot and a human), 2) a pair of animate non-humanoid characters (a fish and a penguin), and 3) a pair of traditionally inanimate objects (a banana and a toilet). The last category was included to reflect traditionally inanimate interactive agents studied in HRI such as a donation box\cite{kim2014effect} or a moving desk \cite{kim2021haunted}. The peach, cat, red panda, and shark were removed to reduce duplication with the banana, penguin, and fish, and the bacteria character was also omitted as we did not believe it would be as popular an interactive agent. Secondly, we modified our prompt, because the participants reported a lack of motivation to initiate interactions when given the original open-ended task. The details are described in Sec. \ref{subsection: study design}. Finally, we added additional rules to the teleoperator guidelines based on the scenarios observed; the full guide is included in the Appendix.
% Emblematic gestures allow people to express a verbal meaning without speech~\cite{ottenheimer2018anthropology, matsumoto2013emblematic}.

\subsection{Study Design}
\label{subsection: study design}
Following our pilot study, we conducted a WoZ study with 20 adults (gender: 9/10 women/men, 1 prefer not to disclose; age: $28.5\pm12.98$) participants interacting with 6 different virtual characters for 1 minute each. The study was approved by the university ethics board. 
% Six characters (Fig. \ref{fig:character}) were chosen to cover a range of morphologies: \textcolor{teal}{TODO: move to pilot study} 1) a pair of humanoid characters (a robot and a human), 2) a pair of animate non-humanoid characters (a fish and a penguin), and 3) a pair of traditionally inanimate objects (a banana and a toilet).

We changed our task prompt to ``Your goal is to test what the character can and cannot do physically, emotionally, and socially" to encourage participants to initiate interaction. This prompt was built upon observed behaviors in the pilot study: posture, proxemics, and physical contact as \emph{``physical behaviors"}, affect displays as \emph{``emotional behaviors"}, and emblematic gestures as \emph{``social behaviors"}. The on-site researcher verbally provided each participant with two non-verbal interactive behaviors as examples, picked randomly from the following list:
\begin{itemize}
    \item waving to others indicates you are greeting someone
    \item nodding to someone at the other end of the hallway to show you noticed them from afar
    \item passing someone your smartphone to show you are trying to share something with them
    \item you can test the characters' emotional capability by testing if they can tell and react to you when you are sad/happy/angry 
    \item pretending to have physical contact with the characters
\end{itemize} 

  In the pre-study briefing, we obtained participants' consent to be video recorded, their permission to release the fully anonymized video data for research purposes, and the identifiable video data for conference presentation. We suspected that people might act differently towards a teleoperated agent from a fully autonomous one, therefore participants were told that all characters were fully autonomous. Participants were also told that the agents were not programmed to process any audio data. Participants were nonetheless allowed to make sounds and speak if they thought this would help them communicate more effectively. During each study session, the participant was instructed to have a 1-minute interaction with each of the 6 virtual characters, following the prompt above. To alleviate the order effect, we randomized the interaction order of the virtual characters, and the researcher turned away from the participant to prevent the participant from feeling observed. At the end of the study, participants were debriefed that the agents were teleoperated, and were asked if they wanted to revise their consent for data release after being debriefed. 

% \textcolor{teal}{We suspected that people might act differently towards a teleoperated agent from a fully autonomous one. Hence, the deception was introduced as we wanted all participants to have a unified perception of agents' control. One ultimate goal of this study is to build a highly socially intelligent agent, however, we don't think existing autonomous agents could deliver believable nonverbal interaction for one minute. Thus, we choose to teleoperate the agents while using deception to achieve our goal.} 

%  During the 1-minute interaction, the researcher in the same room turned away from the participant to let the participant have a comfortable interactive environment without feeling evaluated. The researcher timed each free interaction and instructed the participant at the end of the interaction to fill out a short survey consisting of the Godspeed Questionnaire Series ~\cite{bartneck2009measurement}  Animacy and Likeability scales. In addition, the participants provided a rating of the overall interaction quality and answered two interview questions from the researcher. After finishing all six interaction sessions, participants were asked to fill out a survey of demographic questions and answer a few more interview questions regarding the overall study experience. Further details are provided in Sec. \ref{sec: questionnaire}.

\subsection{Data Collection and Analysis}

We used ELAN~\cite{ELAN} by the Max Planck Institute to annotate our video data. Each video was processed by at least two annotators. The primary annotator segmented and annotated the video, and the secondary annotator went over the existing annotation and took notes of potential disagreement. We adopted the discuss-until-consensus method that is often used in qualitative studies \cite{harding2013analysing}. A third annotator (the teleoperator) was consulted to break disagreements.
% \textcolor{teal}{As most of our annotations were free text annotations, we did not compute inter-rater reliability. Instead, we adopted the discuss-until-consensus method that 

\subsubsection{Segmentation} The primary annotator segmented the entire recorded video into interactive behavior segments. One interactive behavior segment is defined as a complete action-reaction (response) pair, containing (a) an action initiated by either the virtual agent or the participant, and (b) its corresponding response by the other interactor.

\subsubsection{Segment classification} Every segment was classified into one of: physical (posture, proxemics or physical contact), emotional (affect display), or social (emblematic) behaviors. The annotator first determined whether the behavior had a straightforward and interpretable verbal meaning, and if so classified it as a \emph{social behavior}. If not, the annotator examined whether there was a clear emotion expressed via the behavior, and if so, tagged it as an \emph{emotional behavior}. Finally, if the behavior did not fall into either social or emotional behavior categories, the annotator classified it as a \emph{physical behavior}. Thus, this last category contained postural, proxemic, and physical contact behaviors, along with any other behaviors that did not fall into the other two categories. Our method provides us with a hierarchy of behavior understanding, where social behaviors may, but are not required to, be composed of emotional displays, which in turn can be comprised of physical actions. In this way, classifications might not be mutually exclusive from one other. For example, when participants tried to attack the agents with violent actions, they sometimes frowned hard to pretend they were angry. In this case, both emblematic meanings and distinct emotions were annotated.

\subsubsection{Additional labeling} In addition, for each segment, the annotators decided on the following:
 \begin{itemize}
     \item \textit{Initiator of the interaction} (character/participant): the initiator of the interaction
     \item \textit{Character-specific} (yes/no): separates character-agnostic and character-specific behavior by determining if the motive of the behavior is only explainable with the existence of the character's unique features 
     \item \textit{Description of physical actions} (free text): short, physical movement descriptions for the instance
     \item \textit{Emotion label} (free text): the emotion participants tried to express during an interaction instance, if any
     \item \textit{Social meaning} (free text): the social meaning behind an interaction instance, if any
     \item \textit{Response type} (reacting/mirroring/no response): the type of the response from the interaction recipient
     \item \textit{Response description} (free text): short physical/emotional/social descriptions on the interaction recipient's response action, if any
 \end{itemize} 
 The full set of annotations and anonymized videos can be downloaded from \url{https://rosielab.github.io/react-to-this/}.

\subsection{Thematic Analysis}
After video annotation, we set aside the behaviors labeled as character-specific and proceeded with thematic analysis on the remaining (character-agnostic) behaviors. We conducted the thematic analysis using the physical, emotional, and social grouping paradigm derived from the pilot study to form our human behavior codebook:   
\begin{enumerate}
    \item \textbf{Social:} For all the social behaviors (interaction labeled as social/emblematic), we listed all the social meaning annotations and merged those with similar high-level meanings. For example, participants expressed a disapproving attitude with various behaviors such as shaking the head, shaking the finger, thumb(s) down, and forearms crossed as an "X' figure, etc. Thus we extracted a high-level and abstract meaning \emph{disapproval} to summarize this group of behaviors conveying a similar social meaning.
    \item \textbf{Emotional:} We grouped the freely annotated emotion labels based on Ekman's basic emotions, namely anger, disgust, fear, happiness, sadness, and surprise for all the emotional behaviors. For the freely annotated emotion labels that did not seem to fit into the basic emotion categories, we created a separate category on its own.
    \item \textbf{Physical:} We first decided if participants changed their relative spatial distance to the character (proxemics) or if they intended to have physical contact with the character (physical contact). Then, we separated the physical behaviors excluded from the above 2 categories by the activated body parts into 5 groups (full body, head/face, arm, hand, lower body). We observed sequences of movements involving multiple body parts in one physical interactive behavior. For example, participant 19 (P19) jumped, drew a big circle with their arms, widened their eyes, and opened their mouth quickly and continuously in one interaction segment. We counted each individual action and classified each of them into different physical behavior categories to form a detailed thematic result (see Table \ref{tab: physical}).
\end{enumerate}

For character-specific behaviors, we grouped them based on the character conditions, due to the small number of instances (Table \ref{tab: character_specific}).

\section{Results and Discussion}
% \subsection{RQ1: What do people do to test a character physically, emotionally, and socially?} 

\begin{figure*}
    \centering

    \includegraphics[width = \textwidth]{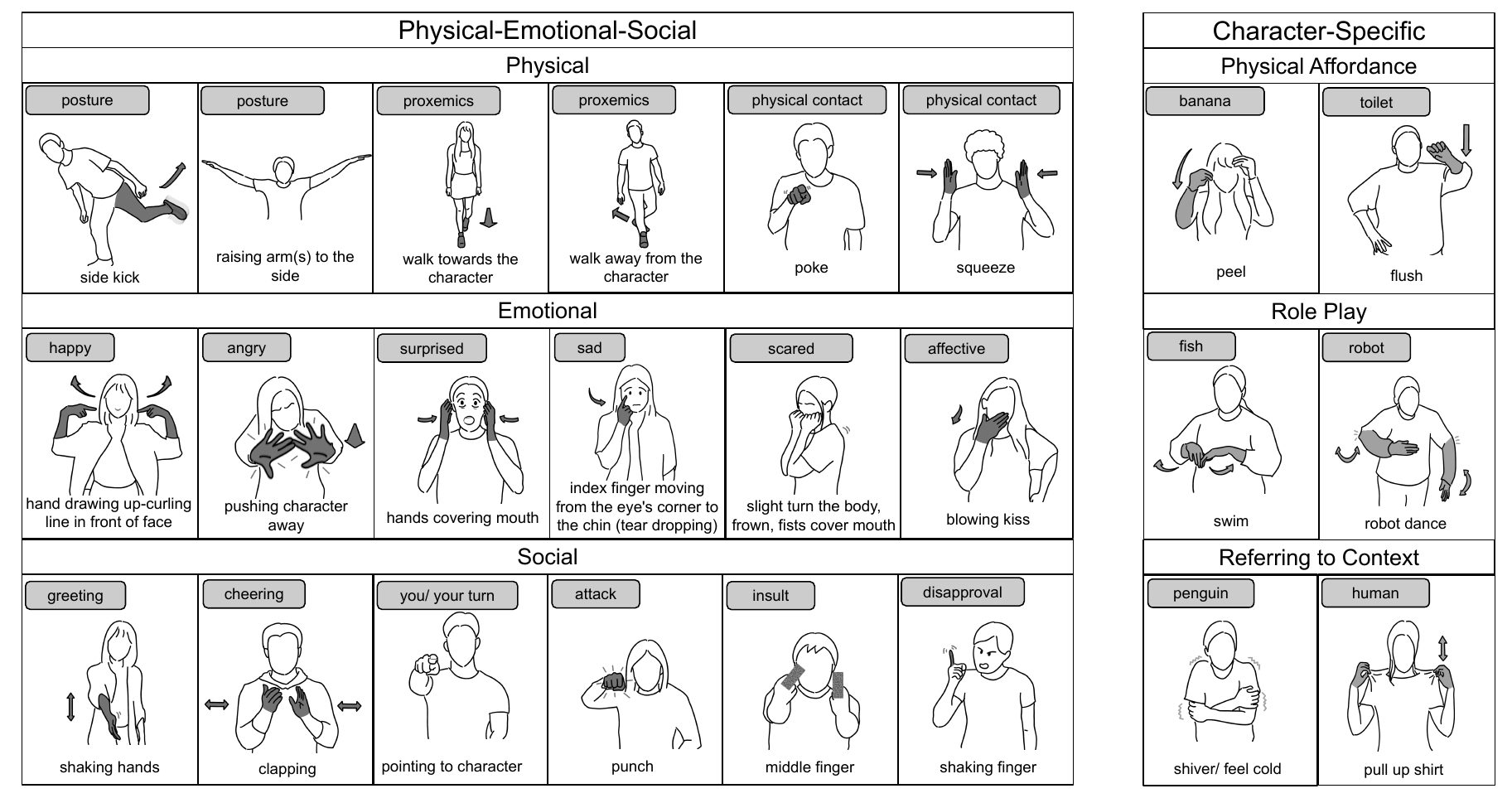}

    \caption{(LEFT) Illustrative examples for physical, emotional, and social nonverbal human behaviors. Physical behaviors included changing posture, proxemic movement, and mock physical contact. (RIGHT) Examples of character-specific interactions related to physical affordance (banana peeling, toilet flushing) role play (fish swimming, robot dancing), and the agent's contextual environment and accessories (shivering from cold,  pulling on shirt).}
 
    \label{fig: social_emotional_physical}
    
\end{figure*}

In total, 1169 interactions were observed during the study, including 1111 interactions initiated by participants and 58 interactions initiated by the agent. From the 1111 interactions started by the participants, we obtained 999 character-agnostic (424 physical instances, 162 emotional instances, 413 social instances), and 112 character-specific behaviors. In this section, we provide results and discussions for each category of behaviors. In addition, we discuss two other behavioral patterns—mimicry and compound behaviors—that emerged from the analysis.

\emph{Physical Behaviors} are presented in Fig. 3 and Table \ref{tab: physical}, and include posture changes, proxemics and physical contact. Firstly, it appeared that posture changes were related to actuation testing, i.e. participants initiated some interactive behaviors (e.g. \emph{leaning} or \emph{raising arms}) to see if the character was able to follow and replicate. Secondly, in some interactions, it appeared characters were not expected to copy the participant's exact movement. Rather, participants were checking whether it could track them (e.g. \emph{walking to the left/right}). We believe this points to a class of behaviors for sensor testing. For example, P113 put their hoodie in front of the robot character to block its view. Finally, in some cases, participants expected physical feedback as a result of their pretend physical contact. Actions involving physical contact contributed to these types of behaviors by testing physics-based responses, such as lifting, pushing, and poking. These are reminiscent of the mild robot abuse behaviors found in \cite{kanda2020}, although we did not necessarily observe aggression (see, however, the \emph{attacking} behavior in Table \ref{tab: social}).

   % \item \textbf{Role-Playing.} Character specific interactions comprised the bulk of behaviors falling into our ``role-playing" category. These included, for example, swimming and puffing out cheeks for the fish, or waddling for the penguin. It is possible that an element of imaginative or pretend play explained some of these behaviors, projecting the participant into the character's world and movements.

\emph{Emotional Behaviors} were conveyed through facial expressions and/or body movements. We found two interesting outcomes: 1) Exaggeration. Participants tried to express sadness by finger-drawing a downward curve in front of their face, moving index fingers from the eye's corner to the chin to depict tear dropping, and rubbing fists around their cheeks to pretend to be crying hard. The occurrence of such pantomime actions might indicate a low expectation of the intelligent virtual characters' capability in emotion recognition. 2) Diversity. We found that one emotion category could comprise multiple unique behaviors. While not surprising based on the psychology literature \cite{barrett2019emotional}, these actions could serve as alternate targets for recognition systems, i.e. instead of attempting to recognize ``sadness", which is an overarching concept comprised of heterogeneous actions, computer vision researchers could aim to recognize ``rubbing eyes, pretending to cry", ``lowered lip corners" and so on. The full list of 34 unique emotional behaviors and 9 categories can be found in Table \ref{tab: emotional}.

\emph{Social Behaviors} were those with clearly identifiable verbal meanings such as greeting, cheering, attacking, insulting, etc. The detailed explanation can be found in Sec. \ref{subsubsec: behavioral categories} under \textit{emblematic gesture}. We observed 82 unique behaviors and derived 42 different social behavior categories from the annotations (Table \ref{tab: social}). Two main observations were: 1) Culture-based behaviors. In addition to well-recognized behaviors such as thumbs up for ``good job", there were also culture-specific behaviors such as namaste (pressing hands together, and fingers pointing upwards). This highlights the importance of investigating culture-specific social signals. 2) Diversity. Similar to the observed emotional behaviors, we found that different behaviors might share similar social meanings. For instance, punching, kicking, and firing handguns can all be used to express one's aggression toward others. 

% \subsection{RQ2: How do these behaviors change depending on the character’s morphology?}
% \label{subsec:results_morphology}
% \onecolumn
% \input{Tables/full_table}
% \twocolumn

% \begin{figure}
%     \centering
%     \includegraphics[width = 0.5 \textwidth]{Figures/character specific_small.png}
%     \caption{Examples for character-specific interactions, related to physical affordance (banana, toilet, human), and roleplay (fish, robot, penguin).}
%     \label{fig: character-specific}
% \end{figure}
\begin{table}[t]

\begin{tabular}{p{1.5cm}|p{6cm}}
    % \hline
    % \multicolumn{2}{c}{\textbf{Character-Specific Behaviors}} \\
    \hline
    \textbf{Character} & \textbf{Nonverbal Character-Specific Behaviors} \\
    \hline
 banana & grabbing; peeling; eating; rotating command by pointing down from above head; bending/breaking; placing the character on hand \\ 
    \hline
    fish & puffing out cheeks; swimming; flipping hands; rolling over command; juggling command; bouncing command\\ 
    \hline 
    toilet & opening and closing lid; flushing; pulling toilet paper; sitting on \\  
    \hline 
    penguin &  waddling; complimenting the scarf; taking off the scarf; shivering (pretending to be cold)\\
    \hline
    robot & robotic arm movement/robot dance;
flapping ears\\
    \hline    
    human & touching hair (e.g. twirling, combing, wearing); complimenting hair; flexing arm; tugging sleeve; grabbing shirt\\
    \hline 
\end{tabular}

\caption{Character specific interactive behaviors. }
  
\label{tab: character_specific}
% \label{tab: character_specific}
% {\lipsum[2] 
% \par
% Table~\ref{tab: character_specific} }
\end{table}
\begin{table}

\begin{tabular}{p{1.5cm}|p{6cm}}
    % \hline
    % \multicolumn{2}{c}{\textbf{Character-Specific Behaviors}} \\
    \hline
    \textbf{Category} & \textbf{Mimicked Behaviors} \\
    \hline
    Physical (20) & swinging (5); raise hands (2); wiggling (2); hand covering mouth (1); shrug (1); wave hands (1); shake head; tilt head (1); lean to the side (1); bend body (1); reaching out arms (1)\\ 
    \hline
    Emotional (5) & smile -- happy (2); hug -- affectionate (2); pouting mouth -- sadness (1); shocked face, raise two hands, lean back -- surprised (1); frowning -- angry (1)\\ 
    \hline 
    Social (16) & shrug (7); ok gesture (6); thumbs up (1); boxing (1); shaking head (1)\\  
    \hline 
\end{tabular}

\caption{Nonverbal behaviors that were initiated by the agent and mimicked by the participants.}

\label{tab: mimicry}
% \label{tab: character_specific}
% {\lipsum[2] 
% \par
% Table~\ref{tab: character_specific} }
\end{table}

\emph{Character-Specific Behaviors} (Fig. \ref{fig: social_emotional_physical} and Table \ref{tab: character_specific}) were defined as the motive of the behaviors are only explainable with the existence of the character's unique features during our annotation process. For example, ``flushing the toilet" and ``pulling the toilet paper" behaviors were only observed during interaction with the toilet, and ``peeling" and ``eating" actions were only observed during the banana interaction session. This might suggest that the physical traits of the character may afford specific physical interactions, and may need to be considered when designing a robot (Fig.~\ref{fig: social_emotional_physical}, right-top). Next, we observed participants ``waddling" with the penguin, ``swimming" with the fish, and ``robot dance" with the robot  (Fig.~\ref{fig: social_emotional_physical}, right-middle). It appeared that participants engaged in role-play with some characters, similar to informal ``play" interactions noted by Dautenhahn~\cite{dautenhahn2002embodied}. Imagining behaviors related to the agent's character may help to predict these types of behaviors (e.g., if designing a robot lion, consider that humans may engage in pantomime roaring). Finally, participants made reference to the imagined environment such as ``shivering" from cold with the penguin, and agent accessories, such as complimenting a shirt or scarf (Fig.~\ref{fig: social_emotional_physical}, right-bottom). This could suggest that humans may be testing the environmental awareness of the agent, ``pointing and referring to areas of and things in it" \cite{ijaz2011}, suggested to be a component of agent believability in video game contexts. Designers may need to consider this when adding accessories to their character.

\begin{table*}[t]
\vspace{2.5mm}
\centering
\begin{tabular}{p{4.5cm}|p{12.4cm}}
    % \hline
    %     \multicolumn{2}{c}{\textbf{Physical Behaviors}}\\
    \hline
    \textbf{Category (Occurrence count)} & \textbf{Nonverbal Physical Behaviors} \\
    \hline
        posture -- full body (136)& turning (36); tilting/leaning body (29); bending the torso sideways (27); jumping (16); body forward or backward (8);  spinning (4); swaying/swing (4); bending the collapsed body pose (4); walking/running/jogging (in place) (3); wiggling (1); kneeling on the ground (1);  jumping jack (1); stretch out legs (1); rotating upper body (1)\\ 
    \hline
    
    posture -- head/face (118)& tilting head (23); open/closed mouth (17); pouting mouth (12); nodding (8); raise eyebrows (7); frowning (6), winking (6); shaking head (5); looking at some direction (5); stick out tongue (5); staring (5); squinting (4); turning head to the side (4); crooked mouth (3); closing one eye (2); rolling eyes (2); sucken cheeks (2); stick out head (1); lips touching nose (1)\\ 
    \hline
    
    posture --  arm (68)  & raising arm(s) (17); stretch arm(s) out (15); wave arms/hands (13);  arms/hands drawing a circle (9); crossed arms (7); arms/hands flapping (2), flipping and rotating wrist (2); open arms (1);  rotating forearm(s) around the elbow(s) (1); bending arm (1)\\ %(8) [9] (77) 
    \hline
    
    posture -- hand (42) & hand(s) touching other body part(s) (23); scratching other body parts (5); clap (3); moving fingers (3); palms together (2); raising hand(s) (2); putting on hood (1); flicking hand (1); hand clasping (1); showing finger(s) (1)\\ 

    \hline
    posture -- lower body (21)& squatting (9); lifting/raising leg(s) up (7); side kick (1); lifting leg(s) to the side (1); stretching out legs (1); standing on toes (1); shaking knees (1)\\ 
    
    \hline  
    proxemics (37) &  walk to the left/right (13); walking away from the character (9); running(5); walking toward the character(4); stepping forward/backward/to the side (4); making big steps(1); walking around(1)\\ 
    
    \hline
    physical contact (13) & push character with hands (4); poking with index finger(s), squeeze character by pinching index finger and thumb (3); grab the character (3); pick up gesture with both hands, put aside (1); lift character up by grabbing and lifting motion (1); squeeze character with palms (1)\\
    \hline      

\end{tabular}

\caption{Physical Behavior Codebook: 73 nonverbal human behaviors initiated by the participants to test the agent's physical abilities, grouped into 7 categories. ";" separates different behaviors.}

\label{tab: physical}
\end{table*}

\begin{table*}[t]
\centering
\begin{tabular}{p{4.5cm}|p{12.4cm}}
    % \hline
    % \multicolumn{2}{c}{\textbf{Emotional Behaviors}} \\
    \hline
    \textbf{Category (Occurrence count)} & \textbf{Nonverbal Emotional Behaviors} \\
    \hline
    angry, annoyed, sullen, stern, aggressive, menacing, unsatisfied (45) &   frown(23); pouting mouth (10); stare(6); shake head (3); shaking finger (3); pushing character away (2)\\ 
    
    \hline
    affectionate (44) & heart gestures (fingers, hands, arms) (22); hugging (15); caressing (1); petting (2); blowing kiss (2);  kissing (1); hands overlap, rest hands on chest (aw gesture) (1)\\ 
    
    \hline
    happy (42) & smile (25); laugh (11); hand drawing an up-curved line in front of face (3); pulling the corners of mouth up (2); giggle (1) \\
    
    \hline 
    surprised, shocked (23) & open mouth (11); \textbf{widen eyes, mouth open} (6); widen eyes/raised eyebrows (4); hands cover mouth (1); fingers spread out around eyes (1)\\ 
    
    \hline
    sad (16)& pouted mouth (7); rubbing eyes (pretending to cry) (3); index finger moving from the eye's corner to the chin (tear dropping) (3); pulling the corner of mouth down with fingers (2);  bowed head (1)\\ 
    
    \hline    
    tired (5) &	\textbf{yawning (3); stretching (1); sighing (1)}\\ 
    
    \hline    
    scared (1) & \textbf{slightly turning the body, frowning, fists covering mouth} (1)\\
    
    \hline 
    shy (1) & \textbf{hands on face, turning away} (1)\\ 
    
    \hline 
    contempt (1) & 	\textbf{turning head to the side, looking down, chin up, side eye} (1)\\ 
    \hline

\end{tabular}

\caption{Emotional Behavior Codebook: 34 nonverbal human behaviors initiated by the participants to test the agent's emotional abilities, grouped into 9 categories. ',' splits one behavior into smaller units to increase the clarity of the action descriptions. ";" separates different behaviors. Compound behaviors are labeled in bold.}
    
\label{tab: emotional}
\end{table*}

\begin{table*}
\vspace{2.5mm}
\centering
\begin{tabular}{p{4.5cm}|p{12.4cm}}
    % \hline
    %     \multicolumn{2}{c}{\textbf{Social Behaviors}} \\
    \hline
    \textbf{Category (Occurrence count)} & \textbf{Nonverbal Social Behaviors} \\

    \hline
    greeting (125) & waving hand (107); shaking hands (5); bow (5); salute (3); raise one hand, moving fingers (1); chin up and down, raise eyebrows quickly (1); two-finger salute (1); upwards nod (1); palms together, fingers facing upward (1) \\ % [8]
    \hline 
    good job (42) &	thumb(s) up (42) \\    %[1]
    \hline   
    approval (29) & nodding (18); ok gesture (11)\\ 
    \hline 
    disapproval (26)	& shaking head (10); shaking finger (5); frowning (5); thumb(s) down (5); forearms crossed as "X" (1)\\ %[2]
    \hline
    you/ your turn (19) & pointing to character (19)\\ %[1]
    \hline 
    dance(19)& causal dancing (11); dab (4); whip (3); nae nae (1)\\    
    \hline
    attacking (19) & boxing/punch (12); fire handgun (3); slap (2); kick (1); hit oneself (1)\\ % [5] 
    \hline
    entertaining (12) & peekaboo (5); making face (3); tada (2); wiggling fingers behind the head (1); flamingo pose (1)\\    % [4]
    \hline 
    I don't know (11)& shrug (11)\\ %[1]
    \hline 
    cheering (11) &	clapping (7); high five (4)\\ 
    \hline 
    deictic(11) & point to some direction for the character to look at/follow (11)\\
    \hline
    questioning (11) & shrug (7); tilt head (2); raise eyebrows, 'asking' face (1); horizontally wave the hand (1)  \\
    \hline
    instruct to copy the exact behavior (10) &	perform certain action and point to the character to instruct the character to replicate the same action (10)\\
    %[2]        
   
    \hline    
    I (8) &	pointing to themselves (5); hand(s) rest on chest (3)\\ %[2]
    \hline 
    come closer (7) & pull hands to oneself (7)\\ %[1]

    \hline 
    thinking (7) & \textbf{index finger over mouth, serious face} (2); fist under chin (2); \textbf{crossed arms, bite lips, nod} (1); hand in chin (1); \textbf{arm crossed, serious face, tilt head} (1)\\ %[1]
    \hline 
    insult (7)	& middle finger (7)\\ %[1]
    
    \hline 
    sleep (7) & \textbf{closed eyes, tilting head, rest head on hands} (7)\\ %[2]
    \hline 
    goodbye (5) &	\textbf{walk away/turn back, waving hand} (5)\\
    \hline
    come with me (4) & \textbf{pull hands quickly towards oneself} (3); point to the back (1) \\   %[1] 
    \hline
    searching (4) &	hands over eyes (4) \\ %[1]
     \hline 
    re-draw attention (4) & wave hand (when characters are facing to the side) (2); finger snapping (1); turn180 degree, then suddenly turn back (1)\\ %[1]   
    \hline 
    peace (3)	& victory gesture (3)\\ %[1] 
    \hline 
    look cute/pretty (2)& hands under chin (aw face)/hands under chin(bare teeth, smile) (2)\\ %[1]
    \hline 
    taunting (2) & \textbf{point at the character, leaning back, laughing} (1); raising both hands, pointing to the character (1)\\
    \hline 
    holding hand (2) & \textbf{reaching out one arm, palm facing up, pointing to the reached-out hand} (1); hold both hands(1)\\ %[1]
    \hline
    interact with a smartphone (2) & pull out a smartphone, pretend to take a picture of the character (1); show the character the smartphone screen (1)\\ %[2]
    \hline
    talking (2)	& \textbf{hands out moving, pretend to talk} (1); hands out (1)\\ %[1]  
    \hline 
    cut it off/stop it (1) & whip hands (1)\\ %[1]    
    \hline
    identity revealing (1) &  pulling down and up the hood on clothes (1)\\ %[2]
    \hline
    eyes on you (1)	& pointing fingers to their own eyes then to the character(1)\\ %[1]
    \hline 
    broken heart (1) & hands making a heart gesture and separate hands (1)\\ %[1]
    \hline
    sick (1) &	sneeze (1)\\ % [1]
    \hline
    call me (1) & hand gesture as a phone, rest the hand next to the ear (1)\\ %[1]

    \hline
    whatever (1) & shrug (1)\\
    \hline
    whispering (1) & \textbf{hand(s) closed to mouth, moving lips} (1)\\ %[1]
    \hline 
    listening (1) &	\textbf{turning body 90 degrees, putting hand close to the ear} (1)\\ %[1]
    \hline 
    numbers (1) & using fingers to indicate some number (1)\\ %[1]
    \hline 
    reading (1) & look at one palm (1)\\ %[1]
    \hline
    write (1) & index finger of one hand hovering over another hand (1)\\ %[1]
    \hline
    comfort/calm (1) & put two hands up, smile (1)\\
    \hline
    so-so (1) & wave the hand horizontally (1)\\
    \hline

\end{tabular}

\caption{Social Behavior Codebook: 82 nonverbal human behaviors initiated by the participants to test the agent's social abilities, grouped into 42 categories. ',' splits one behavior into smaller units to increase the clarity of the behavior descriptions. ";" separates different behaviors. Compound behaviors are labeled in bold.}
  
\label{tab: social}
\end{table*}

\emph{Mimicry} \cite{chartrand1999chameleon} was also observed in this study. From the 57 character-initiated behaviors, we observed that 17 participants mimicked the characters’ behavior or derived the next interactive behavior from it (Table \ref{tab: mimicry}). Although the teleoperator would not initiate interaction on purpose, the presence of delay can interrupt interaction flow; for example, participants might treat a delayed response from an agent as the start of a new interaction. In some of these cases, we observed mimicry behaviors from participants.

It appeared that the mimicry was at times involuntary and unconscious. For example, participant P101 claimed that because he saw that the human character had hair, they actively chose to test if the character was able to groom her hair as he could. However, the recorded video showed that the character touched her hair first, then the participant immediately performed the same action. In another example, P103 reported that after he knocked himself on the head, the banana character appeared to fall asleep (pretending to pass out after being hit), so the participant tried to sleep to examine if the banana actually understood this action. When participants ran out of interaction ideas, consciously mimicking the character's response was one of the methods they used to create more interactive behaviors. As a cautionary guideline for designers, this could suggest that robots may need to recognize all the behaviors that it expresses.

\begin{figure}
    % \centering
    \includegraphics[width = 0.49\textwidth]{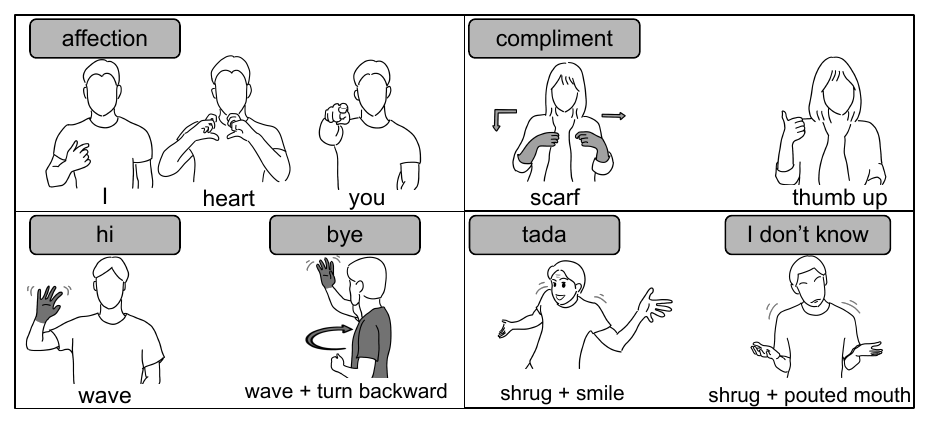}

    \caption{Compound Behavior Examples: behaviors consisting of sequential actions (top) and behaviors consisting of multimodal actions (bottom).}
    
    \label{fig: sequential_multimodal}

\end{figure}

\emph{Compound Behaviors} consisting of sequential and multimodal actions were also observed. For example, (Fig. \ref{fig: sequential_multimodal}, top) P121 drew the shape of the scarf first, then made a ``thumbs up" gesture to give a compliment. Thus, being able to segment action sequences and understand them correctly is crucial for the virtual character to respond appropriately. As an example of a multimodal social signal (Fig. \ref{fig: sequential_multimodal} bottom), the ``shrug" action was seen as a signal of ``I don't know/understand" when it was in the company of raised eyebrows, and was perceived as a sign of ``tada!" when the participants held the shrug for a long time and displayed a surprised-happy face. Sequential and multimodal actions make segmenting and recognizing behavior a challenging problem for future work.

\section{Limitations and Future Work}
Our research was conducted in a North American post-secondary institution. Thus, our findings may not be representative of the general public from different cultural backgrounds. Future research may target a larger and more diverse sample (including people who are of different ages, who are neurodiverse, or who are from varied cultural groups). Also, virtual agents allowed for flexible testing of various morphologies, but physical interaction with real robots may differ. For example, Kanda et al. found that children try to obstruct the path of a navigating robot \cite{kanda2020}. We also did not investigate auditory nonverbal behaviors, such as clapping or non-linguistic utterances. Finally, we do not claim that our list of interactive behaviors is exhaustive, but rather contributes to the set of nonverbal human behaviors under scrutiny by HRI researchers. The next step of this research involves creating algorithms to recognize these behaviors, and investigating the effect of various agent reactions to them. We hope that this set of behaviors can be applied to robots and interactive agents deployed in various settings (e.g. theme parks, education, video games) to increase believability in the future.

% At the time of writing this paper, we were not aware of any existing dataset consisting of over 137 actions/gestures. Although we are still far from creating a complete handbook with exhausted action/gesture taxonomies.

\section{Conclusion}
In this study, we discovered 188 unique actions and 51 socio-emotional behavior categories among 1169 non-verbal interactions between participants and 6 virtual characters, contributing to the list of target classes for visual gesture recognition algorithms. With a bottom-up analysis method, we created a rich and diverse behavior codebook to guide designers and programmers of interactive agents/robots. The 188 actions and corresponding meanings also could help provide a list of classes for machine learning gesture recognition algorithms to target. The set of abundant interactive behaviors in our codebook can be applied to interactive agents deployed in various settings (e.g. video games, theme park, education) in the future.

\bibliographystyle{IEEEtran}
\bibliography{custom.bib}
\appendix
\subsection{Teleoperator Guidelines}
The teleoperator followed the following guidelines during interactions with the participants. For each participant action (left of the arrow), the virtual character should show a corresponding response (right of the arrow):
\begin{itemize}
    \item Greeting $\rightarrow$ mirroring
    \item Pointing directions $\rightarrow$ turning/leaning/looking/pointing at/towards the direction
    \item Physical contact $\rightarrow$ pretend the physical contact is happening
    \item Show aggressiveness/ disgust face $\rightarrow$ show hurt/angry
    \item For an expression that you cannot do with the character $\rightarrow$ use the character to make the expression ‘I cannot do that’
    \item If no interaction performed $\rightarrow$ looking around (idle mode)
\end{itemize}

% \subsection{Additional Trends}
% % \begin{wrapfigure}{r}{0.5\textwidth}
%   \begin{figure}[hbt!]
%         % \centering
%         \includegraphics[width = 0.4\textwidth]
%         {CHI2024/Figures/interaction_boxplot.png}
%         \includegraphics[width = 0.4\textwidth]{CHI2024/Figures/awareness_boxplot.png}
%         \caption{Interaction quality rating for all characters}
%         \label{fig: interaction quality_awereness}
%   \end{figure}  
% % \end{wrapfigure}

% \begin{figure}[hbt!]
%     % \centering
%     \includegraphics[width = 0.7\textwidth]{ubicomp2024/Figures/likeability_animacy.png}
%     \caption{Correlations between Godspeed Likeability Scale and Animacy Scale for all characters}
%     \label{fig: likability_animacy}
% \end{figure}

% \begin{figure}[hbt!]
%     % \centering
%     \includegraphics[width = 0.4\textwidth]{CHI2024/Figures/Interaction_ranking.png}
%     \includegraphics[width = 0.4\textwidth]
%     {CHI2024/Figures/Cuteness_ranking.png}
%     \caption{Character Ranking based on interaction quality and perceived cuteness level}
%     \label{fig: ranking}
% \end{figure}

% \input{Tables/physical_ly}
% \input{Tables/emotional_table}
% \input{Tables/social_table}

\end{document}